\documentclass{epl}

\usepackage{graphics}
\usepackage{overpic}

\newcommand{\beq}{\begin{equation}}
\newcommand{\eeq}{\end{equation}}
\newcommand{\bea}{\begin{eqnarray}}
\newcommand{\eea}{\end{eqnarray}}

\newcommand{\rmd}{{\rm d}}
\newcommand{\rmi}{{\rm i}}

\title{Metallic Continuum Quantum Ferromagnets at Finite Temperature}
\author{M.~B.~Silva~Neto \and A.~H.~Castro~Neto}
\institute{Department of Physics, Boston University, Boston, MA 02215}


\begin{document}

\maketitle

\begin{abstract}
We study via renormalization group (RG) and large $N$ methods the problem of 
continuum SU(N) quantum Heisenberg ferromagnets (QHF) coupled to gapless 
electrons. We establish the phase diagram of the dissipative 
problem and investigate the changes in the Curie temperature, 
magnetization, and magnetic correlation length due to dissipation
and both thermal and quantum fluctuations. We show that the interplay
between the topological term (Berry's phase) and dissipation leads to 
non-trivial effects for the finite temperature critical behavior.
\end{abstract}

\section{Introduction}

The subject of quantum ferromagnetism in metallic and semiconducting
environments has recently attracted a lot of interest due to its relevance 
to a broad class of physical systems such as heavy fermion compounds, for 
instance URu$_{2-x}$Re$_{x}$Si$_{2}$ \cite{Maple} and
Th$_{1-x}$U$_{x}$Cu$_{2}$Si$_{2}$
\cite{Greg}, dilute magnetic semiconductors such as Ga$_{1-x}$Mn$_{x}$As 
\cite{ohno}, ferromagnetic dichalcogenides CeTe$_{2}$ \cite{jung}, 
manganites La$_{1-x}$Sr$_{x}$MnO$_{3}$ \cite{manganitas}, and 2D electron 
systems in the quantum Hall regime \cite{quantum-Hall}. These systems 
are characterized by the presence of local moments that couple directly 
to an itinerant electron liquid and can be described in terms of the 
Hamiltonian:
\beq
{\cal H}=-J\sum_{\langle i,j\rangle}{\bf S}_{i}\cdot{\bf S}_{j}+
J_{K}\sum_{i,\sigma,\sigma^{\prime}}{\bf S}_{i}\cdot
c^{\dag}_{i,\sigma}\vec{\sigma}_{\sigma\sigma^{\prime}}c_{i,\sigma^{\prime}}+
{\cal H}_{e},
\label{Hamiltonian}
\eeq
where $J$ is a Heisenberg exchange between localized spins ${\bf S}_{i}$
(total spin $S$), $J_{K}$ is an exchange coupling between localized spins 
and conduction electrons, $c^{\dag}_{i,\sigma}$ ($c_{i,\sigma}$) is the
creation (anihilation) electron operator, $\vec{\sigma}$ are the Pauli 
matrices, and ${\cal H}_{e}$ describes the conduction electrons.

On the theoretical side, the critical properties of the {\it insulating} 
QHF have been studied via RG methods \cite{chakravarty} and it has been 
shown that both the correlation length and the magnetic susceptibility 
exhibit renormalized classical behavior at low temperatures and that no 
non-trivial quantum critical point (QCP) exists, in agreement with Monte 
Carlo data \cite{monte-carlo}. Other formulations of the insulating QHF 
based on O(N) and SU(N) models were used to explain magnetization and 
relaxation rate in incompressible quantum Hall systems \cite{sachdev,girvin}. 
The problem of {\it itinerant} Hubbard ferromagnets, on the other hand, was 
studied by Hertz \cite{hertz} in the context of {\it classical} 
Ginzburg-Landau-Wilson theory. This approach has been criticized recently 
because it does not take into account the physics of soft-modes \cite{belitz}. 

In a metallic system such as the one described by (\ref{Hamiltonian}) 
dissipation occurs because electrons scatter from the localized moments 
and act as a heat bath for the spin dynamics. There are two main sources 
of dissipation: Landau damping as in the case of clean magnets, and 
electronic diffusion in the case of structurally disordered magnets 
\cite{fulde}. We are going to use both momentum shell RG and large 
$N$ analysis to generate the phase diagram of the dissipative problem. 
As we are going to show, a new QCP arises from dissipation. We find, 
moreover, that dissipation is a {\it relevant} perturbation under the 
RG and produces a runaway flow to strong dissipation. Our results from 
the RG analysis agree with the large $N$ solution of the problem. 
Finally, we discuss the effects of dissipation on the finite temperature 
critical properties of the system (\ref{Hamiltonian}). More specifically 
we discuss how the Curie temperature is affected by the different 
types of conduction electron dynamics.

\section{The effective field theory}

Our starting point is the partition function for the system described by
(\ref{Hamiltonian}) (we use units such that $\hbar=k_B=1$):
\beq
Z=\int{\cal D}{\bf n}{\;}\delta({\bf n}^{2}-1)
{\;} e^{-S_0({\bf n})-S_I({\bf n})-S_D({\bf n})},
\label{z}
\eeq
where
\beq
S_0({\bf n})=\int_{0}^{\beta}\rmd\tau\int\rmd^{d}{\bf x}
\left\{M_{0}{\bf A}({\bf n})\cdot\partial_{\tau}{\bf n}+
\frac{\rho^{0}_{s}}{2}(\nabla{\bf n})^{2}\right\},
\label{Eff-Action-n}
\eeq
is the long wavelength Euclidean action of the ferromagnetic O(N) 
non-linear sigma model \cite{sachdev} and $\beta = 1/T$. Here 
${\bf n}$ is a $N$ component vector that represents the local 
magnetization, $\rho^{0}_s = J S^2 a^{2-d}$ is the spin-stiffness 
($a$ is the lattice spacing, and $J$ is the nearest neighbor 
ferromagnetic exchange), and $M_0 = S a^{-d}$ is the magnetization 
density in the ground state. The main difference between 
this action and the Hertz action, usually used for the study of itinerant 
quantum critical phenomena \cite{hertz}, is the presence of the topological 
term (Berry's phase) described by the vector potential ${\bf A}$ of a Dirac 
monopole at the origin of spin space, a signature of the presence of 
localized moments. As we shall see, this term is fundamental for the 
understanding of the problem described by (\ref{Hamiltonian}). Furthermore, 
while (\ref{Eff-Action-n}) describes a problem with short range interactions 
we can also introduce long range interactions that decay like $1/r^{d+\sigma}$ 
by modifying the gradient term in momentum space from $k^{2}$ to $k^{\sigma}$ 
with $\sigma \leq 2$ \cite{fisher}. 

The action $S_I$ in (\ref{z}) contains higher spatial derivative terms 
(with no time derivatives) corresponding to spin-wave scattering. It 
can be shown from a RG analysis that the coupling constant $\lambda$ 
associated with these terms has scaling dimension $d+z-4$ and therefore 
is irrelevant for $d<4-z$, where $z$ is the dynamical exponent
\cite{sachdev}. Near the quantum critical point at $T=0$ we have $z=2$ 
for an insulating magnet and $z=3 (4)$ for a clean (dirty) magnet with 
local interactions \cite{hertz}. This shows that $\lambda$ is irrelevant 
only if $d<2,1,0$, for $z=2,3,4$ respectively, otherwise producing 
corrections to the scaling of physical quantities. For the finite 
temperature $T\neq 0$ critical phenomena, on the other hand, we can set 
$z=0$ and treat the temperature as a free parameter. In this case 
$\lambda$ is irrelevant in $d<4$, and for this reason we 
ignore $S_{I}$. We shall however keep $z$ undetermined 
throughout our calculations in order to compare our results
with other approaches.

The third term in (\ref{z}) contains the coupling to the
dissipative environment and it has the form \cite{fulde}:
\beq
S_D = a^{d}\beta \sum_{n,{\bf k}} 
\eta_0 \frac{|\omega_n|}{k^{\delta}} {\bf n}({\bf k},\omega_n) \cdot 
{\bf n}(-{\bf k},-\omega_n),
\label{s2}
\eeq
where $\omega_n=2\pi n/\beta$, and $\eta_0$ is the coupling to the 
electronic bath. The momentum dependence of the dissipative term depends 
on the electronic dynamics: Landau damping gives $\delta=1$ (clean case)
while diffusion implies $\delta=2$ (dirty case) \cite{fulde}.

In what follows we rewrite the partition function (\ref{z}) in terms of a 
basis of boson coherent states (similar results can be obtained in the O(N) 
representation) \cite{girvin}. We write 
$n_i(\tau,{\bf x})=\bar{z}_{\alpha}(\tau,{\bf x}) 
\sigma^i_{\alpha\beta}z_{\beta}(\tau,{\bf x})$ 
where $z_{\alpha}$ are complex bosonic fields subject to the condition that 
$\sum_{\alpha}\bar{z}_{\alpha} z_{\alpha}=N$. 
In terms of these new fields $S_0$ reads:
\beq
S_0=\frac{1}{g_{0}}
\int_{0}^{\beta}\rmd\tau\int\rmd^{d}{\bf x}
\left\{c_{0} \, \bar{z}\partial_{\tau}z+
\left|(\nabla-\rmi{\cal A})z\right|^{2}
\right\},
\label{Eff-Action-z}
\eeq
where $g_{0}=N/(2\rho_{s}^{0})$ is the bare coupling constant, 
${\cal A}=\rmi\bar{z}_{\alpha}\nabla z_{\alpha}$ 
is a gauge field, and $c_{0}=2M_{0}g_0/N$ is the bare topological 
constant (here the sum over $\alpha=1,..,N$ is implicitly assumed). 
Such formulation gives very good results at low temperatures even
for $N=2$ but fails to reproduce the correct values for the critical 
exponents near $T_{c}$ \cite{katanin}. However it is exact at all
temperatures when $N\rightarrow\infty$. Moreover, as we are going 
to show, our results for the effects of dissipation on the magnetism 
are already observed at the lowest temperatures, thus being robust, 
and should hold at higher temperatures as well, in particular at 
$T_{c}$, irrespective of the type of formalism employed.

While the non-interacting part of the action is quadratic in the $z_{\alpha}$
fields, the dissipative term $S_D$ turns out to be quartic. We can show, 
however, that in the large $N$ limit the leading contribution to the boson 
self-energy is given by \cite{next}:
\begin{eqnarray}
S_{D} \approx a^{d}\beta\langle\eta_{0}\rangle 
\sum_{n,{\bf k}} \frac{|\omega_n|}{k^{\delta}}
\bar{z}(\omega_n,{\bf k}) z(-\omega_n,-{\bf k}) \, ,
\label{seff}
\end{eqnarray}
where $\langle\eta_{0}\rangle=\eta_{0}\langle\bar{z}z\rangle$.
Thus dissipation in the scalar vector field ${\bf n}$ also induces 
dissipation in the bosons. 

Because of the constraint in (\ref{z}) the problem described here is 
non-linear. Introducing a Dirac delta function for the constraint, 
the partition function can be written as
$Z \approx \int{\cal D}\phi{\cal D}\bar{z}{\cal D}z{\cal D}{\cal A} 
\exp\left\{-S+\frac{N}{g_0} \int \rmd\tau \int \rmd^{d}{\bf x}
\phi(\tau,{\bf x})\right\}$, where $\phi(\tau,{\bf x})$ is a 
Lagrange multiplier field, and
\beq
S=\frac{a^{d}\beta}{g_{0}} 
\sum_n \int \frac{\rmd^{d} {\bf k}}{(2 \pi)^d} \chi_0^{-1}({\bf k},\omega_n)
\bar{z}(\omega_n,{\bf k}) z(-\omega_n,-{\bf k}), 
\nonumber
\eeq
with 
\beq
\chi_0^{-1}({\bf k},\omega_n) = \left|{\bf k}-{\cal A}\right|^{\sigma} - 
\rmi c_0 \omega_n + \phi + \gamma_{0} \frac{|\omega_n|}{k^{\delta}}.
\eeq
Here we have defined $\gamma_{0}=\langle\eta_{0}\rangle g_0/N$. Because of the 
quadratic form of $S$ the fields $z$ can be integrated out exactly and 
the partition function becomes 
$Z = \int{\cal D}\phi{\cal D}{\cal A} \exp\left\{-N S_{eff}[\phi,{\cal A}]\right\}$
where
\beq
S_{eff}[\phi,{\cal A}] = \frac{1}{g_0} \int_0^{\beta} 
\rmd\tau \int \rmd^{d}{\bf x} \,  \phi(\tau,{\bf x})
+ tr \ln{\chi_0^{-1}(\phi,{\cal A})}.
\eeq
When $N \to \infty$ the partition function is dominated by the saddle 
point where $\langle\phi\rangle=r_0$ is constant and 
$\langle{\cal A}\rangle=0$. The constant $r_0$ is determined by 
the saddle-point equation:
\beq
\frac{1}{g_{0}}=\frac{1}{a^{d}\beta} \sum_{n,{\bf k}}
\frac{1}{k^{\sigma} - \rmi c_0 \omega_n + r_0 + 
\gamma_{0} \frac{|\omega_n|}{k^{\delta}}}{\,} .
\eeq
Notice that $r_0(T) = 1/\xi^{\sigma}(T)$ where $\xi(T)$ is the magnetic
correlation length. After summing over the frequencies we obtain:
\beq
\frac{1}{g_{0}}=\int^{\Lambda}\frac{\rmd^{d}{\bf k}}{(2\pi)^{d}}
\int_{-\Omega^{0}_{k}}^{\Omega^{0}_{k}}\frac{\rmd\omega}{\pi}
\frac{k^{\delta}{\;}\gamma_{0} {\;}\omega{\;}n_{B}(\beta\omega)}
{(k^{\delta}(k^{\sigma}+r_{0}-c_{0}\omega))^{2}+\gamma_{0}^{2}\omega^{2}},
\label{saddle-point-eq}
\eeq
where $n_{B}(z)=(e^{z}-1)^{-1}$ is the Bose occupation number, $\Lambda
\approx 1/a$ is the high energy cut-off (for $d=3$ we choose $(\Lambda a)^3 =
6 \pi^2$ for a cubic lattice so that the volume of the Brillouin zone is
preserved), and $\Omega^{0}_{k}=(\Lambda^{\sigma}/\gamma_{0})k^{\delta}$ 
is the bare frequency cutoff \cite{hertz}.

The RG calculation can be accomplished by splitting the momentum 
integral into two pieces: from $0\leq k\leq\Lambda/b$ and another 
from $\Lambda/b\leq k\leq\Lambda$, with $b = e^{\ell}$. After a 
change of variables, $k=x^{2/\sigma}(\Lambda/b)$ and 
$\omega=y(\Lambda/b)^{z}$, the parameters of the problem are 
rewritten as: 
$r=r_{0}(\Lambda/b)^{-\sigma}$, $c=c_{0}(\Lambda/b)^{z-\sigma}$,
$\gamma=\gamma_{0}(\Lambda/b)^{z-(\delta+\sigma)}$,
$g=g_{0}S_{d}(\Lambda/b)^{d+z-\sigma}/\sigma$, and
$t=g_{0}S_{d}(\Lambda/b)^{d-\sigma}/\sigma\beta$, where 
$S^{-1}_d=2^{d-1}\pi^{d/2}\Gamma(d/2)$. 
In terms of the new variables, the saddle-point equation becomes:
\beq
1=\frac{2 {\cal Z} g}{\pi}
\int_{0}^{1}\rmd x \int_{-\Omega_{k}}^{\Omega_{k}}\rmd y
\frac{x^{2(d+\delta)/\sigma - 1}{\;}\gamma{\;}y{\;}n_{B}(g y/t)}
{(x^{2\delta/\sigma}(x^{2}+r-cy))^{2}+(\gamma y)^{2}},
\eeq
where
\beq
{\cal Z}^{-1}=1-\frac{2 g}{\pi}
\int_{-\Omega_{k}}^{\Omega_{k}}\rmd y
\frac{\gamma{\;}y{\;}n_{B}(g y/t)}
{(1+r-cy)^{2}+(\gamma y)^{2}}
\times{\;}{\ell},
\label{Renormalization-Constant}
\eeq
is a renormalization constant accounting for the contributions of the
high energy modes. From ${\cal Z}$ we obtain the RG equations to 
leading order in $1/N$ \cite{next}:
\bea
\partial_{\ell} {\;} r &=& \sigma {\;} r, \nonumber \\
\partial_{\ell} {\;} c &=& (\sigma-z){\;} c, \nonumber \\
\partial_{\ell} {\;} \gamma &=& (\delta+\sigma-z){\;} \gamma, \nonumber \\
\partial_{\ell} {\;} g &=& -(d+z-\sigma){\;}g+g^{2}{\;}\Phi(r,c,\gamma,g,t), \nonumber \\
\partial_{\ell} {\;} t &=& -(d-\sigma){\;}t+g{\;}t{\;}\Phi(r,c,\gamma,g,t),
\label{rgflow}
\eea
with
\beq
\Phi(r,c,\gamma,g,t)=
\frac{2}{\pi}
\int_{-\Omega_{k}}^{\Omega_{k}}
\rmd y\frac{\gamma{\;}y{\;}n_{B}(g y/t)}
{(1+r-c y)^{2}+(\gamma y)^{2}}.
\label{Phi}
\eeq
For $d \geq 4-z$ we should add to
the above set the equation 
$\partial_{\ell}\lambda=(d+z-4)\lambda+\rho^{0}_{s}/M_{0}$
which gives the flow of $\lambda$ \cite{sachdev}.

We find that there are three fixed points in the $(t,g,\gamma)$
parameter space: $(0,0,0)$ which is the trivial fixed point with 
$r=0$ (that is, with long range order); $(t_*=(d-\sigma)/2,0,0)$ 
which is a unstable fixed point that marks the transition from 
ferromagnetism to paramagnetism; and a line of non-trivial
fixed points at $(0,g_{c}(\gamma),\gamma)$, with $r=0$, that 
separates a fully polarized ferromagnet from a partially 
polarized ferromagnet \cite{next}. Furthermore, we notice 
that the lower critical dimension ($t_{*}>0$) is given by 
$d_{lc}=\sigma$. For short range interactions, $\sigma=2$, 
and order is only possible above two dimensions in accordance 
to the Hohenberg-Mermim-Wagner theorem. For long range 
interactions, $\sigma<2$ and long range order is possible 
for $d \leq 2$ because even long range distortions of the 
ferromagnetic order cost finite energy. 

\begin{figure}
\includegraphics[scale=0.34]{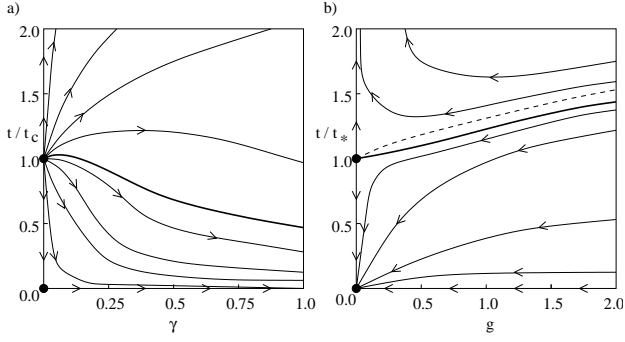}
\caption{RG flow for $d=3$, $\sigma=2$ and $\delta=2$: (a) flow in the
$t \times \gamma$ plane; (b) flow in the $t \times g$ plane for $g<g_{c}$. 
In (b) we compare the cases without (dashed line) and 
with (solid line) dissipation.}
\label{fig1}
\end{figure}

We can now discuss the case of quantum critical phenomena when
$z \neq 0$. For  $\gamma=0$ we recover the results for the insulating 
QHF by setting $z=\sigma$ so that $c$ is unrenormalized under the RG 
while $\gamma$ is found to be a relevant \cite{chakravarty} perturbation. 
In fig.~\ref{fig1}(a) $t_c$ is the solution of $\partial_{\ell} t=0$ for 
$\gamma=0$ and finite $g$, that is, it is the critical temperature of 
the insulating QHF and is depicted as the dashed line in fig.~\ref{fig1}(b). 
For $\gamma\neq 0$, on the other hand, we can choose $z=\delta+\sigma$ so 
that $\gamma$ remains unrenormalized and we obtain $z=3 (4)$ for the 
metallic ferromagnet with local interactions and Landau damping (diffusion). 
In this case, we see that $c$ becomes irrelevant and flows to zero
signaling the suppression of localized moments and making the renormalized
action look very similar to Hertz action for the itinerant case. 
Although this result shows that $c$ is irrelevant under the RG, one 
has to remember that the RG indicates the behavior of the system at 
very low temperatures, deep inside of the ferromagnetic phase if 
$d>\sigma$. However, at finite energy scales (finite temperatures) 
$c$ produces important changes in the behavior of $T_c$ as we explain 
below. Finally, we remark that a complete treatment of the RG for 
$z\neq 0$ should also take into account the corrections from the flow 
of $\lambda$ to the scaling of all the other couplings \cite{sachdev}.

\begin{figure}
\includegraphics[scale=0.31,angle=-90]{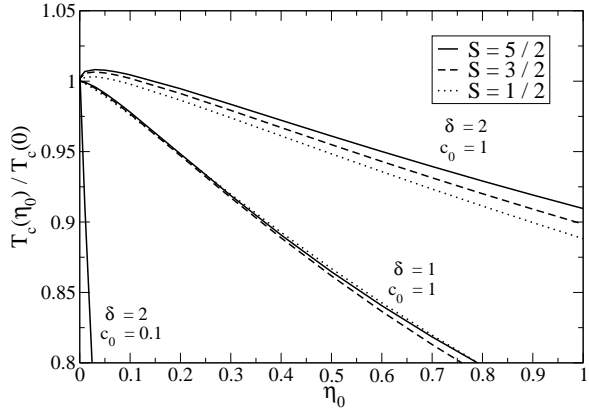}
\caption{Curie temperature $T_{c}(\eta_{0})/T_{c}(\eta_{0}=0)$ as a function 
of dissipation $\eta_{0}$ for different values of $S$ and $\delta$. Note
that $T_{c}$ for the dirty case is always higher than for the clean 
case.}
\label{fig2}
\end{figure}

\begin{figure}
\includegraphics[scale=0.31,angle=-90]{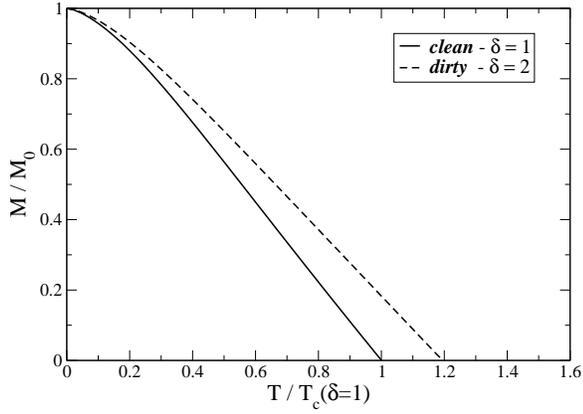}
\caption{Reduced magnetization $M/M_{0}$ as a function of temperature
for $S=5/2$ and $\eta_0=1.0$.} 
\label{fig3}
\end{figure}

\begin{figure}
\includegraphics[scale=0.31,angle=-90]{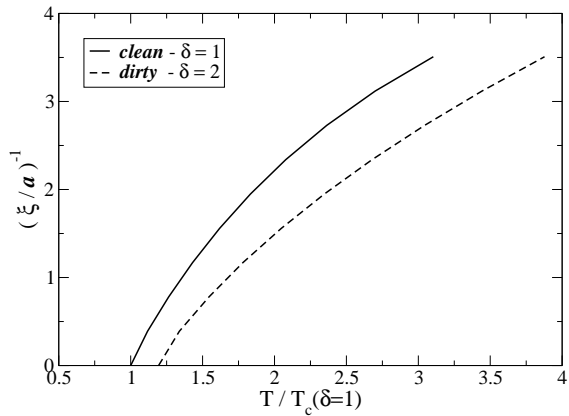}
\caption{Inverse correlation length as a function of temperature for 
$S=5/2$ and $\eta_0=1.0$.}
\label{fig4}  
\end{figure}

In order to understand the critical behavior at finite temperatures 
we turn to the solution of saddle point equation (\ref{saddle-point-eq}). 
The critical line for ferromagnetic order is obtained by imposing that 
$r_{0}=0$, for $\beta=\beta_{c}$. In fig.~\ref{fig2} we show the behavior 
of the Curie temperature as a function of the dissipation parameter $\eta_0$. 
We clearly see that the Curie temperature decreases faster for Landau damping 
than for diffusion. The physical interpretation is that ballistic electrons 
have an infinite mean free path and strongly scatter from the localized 
magnetic  moments, thus introducing a large amount of fluctuations into the 
system. Diffusive electrons, on the other hand, have a finite mean free path 
and are consequently less effective in increasing fluctuations. We have also 
noticed an increase of about $1\%$ on the value of the Curie temperature for 
small enough $\eta_0 \ll 1$ with respect to the case without dissipation. The 
enhancement of the magnetism due to diffusion also seems to be important in 
the 2D MOSFET problem \cite{chamon}. For $c_{0} \rightarrow 0$ we find that 
$T_c$ is rapidly suppressed by the dissipation as shown in fig.~\ref{fig2},
showing the importance of the topological term for the finite temperature
critical phenomena. Remarkably, as $\eta_{0}\rightarrow 1$, we observe an 
enhancement of the magnetism of about $20\%$ with a $T_{c}$ higher in the 
diffusive (dirty) regime relative to the ballistic (clean) case. Real 
systems should then be found in between these two limits. We have 
calculated other physical properties of interest such as the magnetization, 
$M$, and the magnetic correlation length, $\xi$, as a function of 
temperature for the two limiting cases of Landau damping and diffusion 
for a fixed value of $\eta_0$. The results are shown in figs.~\ref{fig3} 
and \ref{fig4}. As we see from fig.~\ref{fig3} the enhancement of the 
magnetism is already present at low enough temperatures where the 
formalism is more adequate. Consequently, such a behavior is expected 
to be observed within other formulations of the problem that captures 
the correct physics at high enough temperatures. 

In summary, we have studied the problem of metallic quantum ferromagnets
at finite temperatures using large $N$ and RG techniques and established the 
phase diagram for this problem.  We found that from the point of view of the 
insulating ferromagnet, $z=\sigma$, any  small amount of dissipation produces 
a runaway RG flow to strong dissipation. We found a line of new QCP 
corresponding to a transition to a not fully polarized ferromagnet at 
$g>g_{c}$, to be investigated elsewhere \cite{next}. We also found that 
the topological and dissipative terms have a rather non-trivial competition 
that, as a consequence, produce a higher Curie temperature for diffusive 
than for ballistic electrons. Our theory provides a general and unified 
framework for the study of both insulating as well as metallic magnets 
(clean or structurally disordered) at vanishing and finite temperatures. 
Applications of our results to experimentally relevant systems will be 
given in future publications.

%
%
%
%
%
%
%
%

\acknowledgments

We acknowledge invaluable discussions with I.~Affleck, K.~Bedell,
C.~Chamon, P.~Kopietz, A.~H.~MacDonald, E. Novais, S.~Sachdev, and 
G.~Stewart. M.~B.~Silva~Neto acknowledges CNPq (Brazil) for financial 
support and Boston University for the hospitality.

\end{document}